# In situ characterization of few-cycle laser pulses in transient absorption spectroscopy

Alexander Blättermann*, Christian Ott, Andreas Kaldun, Thomas Ding, Veit Stooß, Martin Laux, Marc Rebholz, and Thomas Pfeifer

*Max Planck Institute for Nuclear Physics, Saupfercheckweg 1, 69117 Heidelberg, Germany*
*Corresponding author: blaetter@mpi-hd.mpg.de*

Attosecond transient absorption spectroscopy has thus far been lacking the capability to simultaneously characterize the intense laser pulses at work within a time-resolved quantum-dynamics experiment. However, precise knowledge of these pulses is key to extracting quantitative information in strong-field highly nonlinear light-matter interactions. Here, we introduce and experimentally demonstrate an ultrafast metrology tool based on the time-delay-dependent phase shift imprinted on a strong-field driven resonance. Since we analyze the signature of the laser pulse interacting with the absorbing spectroscopy target, the laser pulse duration and intensity are determined *in situ*. As we also show, this approach allows for the quantification of time-dependent bound-state dynamics in one and the same experiment. In the future, such experimental data will facilitate more precise tests of strong-field dynamics theories.

In the transformative field of ultrafast light–matter interaction, characterization of strong-field laser pulses is crucial in order to draw quantitative conclusions from measurement results. Based on the detection of photoelectrons, the attosecond streaking technique provides insight into the system's underlying quantum dynamics on the natural electronic time scale, combined with a powerful characterization of the used strong-field laser pulses [1-3]. Being capable of resolving bound-state quantum dynamics, the all-optical method of attosecond transient absorption spectroscopy (ATAS) [4-12] is an important complement to photoelectron-based methods. In addition to being sensitive to bound-bound transitions, ATAS allows targets to be studied all the way from a natural – weak-field – environment up to the strong-field regime. Using attosecond pulsed extreme-ultraviolet (XUV) light together with (typically) femtosecond near-infrared (NIR) pulses in a pump-probe scheme, dynamical processes leave their fingerprints in the spectrum of the transmitted XUV light. Yet unlike streaking, up to now ATAS lacks the possibility for *in-situ* characterization of the femtosecond NIR laser pulse, which drives and controls the underlying electron dynamics. However, precise knowledge of its shape and intensity would enhance the scope of ATAS concerning quantitative measurements and the control of quantum dynamics in general.

Here, we demonstrate a new scheme for ATAS combining the measurement of time-resolved quantum dynamics with the characterization of the few-cycle driving laser pulse in one single experiment. Our method is based on extracting the time-delay-dependent phase shift [11,13-15] imprinted on a resonance, which is excited by XUV light and dressed by the femtosecond laser pulse that is to be characterized. From this phase shift, we are able to directly obtain the laser pulse duration, as well as its time dependent intensity. Since we analyze the signature of the laser pulse interacting with the XUV-excited spectroscopy target, the pulse characteristics are determined *in situ*. The precise knowledge of laser pulse characteristics in ATAS will enable a more detailed analysis of bound-state dynamics in strong and short laser

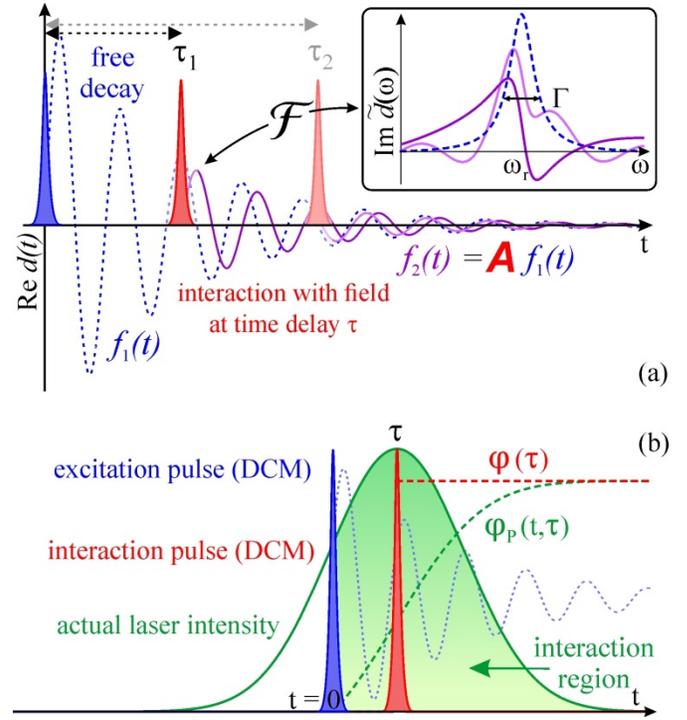

Fig. 1. Dipole control model. (a) Perturbations to the field-free evolution of the dipole response (dashed blue) are treated as instantaneous modifications of the amplitude and phase expressed by the complex factor A. The perturbed decay (dark purple, light purple) is shown for two time-delays ($\tau_1, \tau_2$) leading to very different spectral line shapes (inset). (b) Treatment of a ponderomotive shift in the dipole control model. Instead of a continuously modified dipole phase $\varphi_P(t)$, the effect of the dressing field is approximated by a single phase step at time $\tau$, the peak of the intense laser pulse envelope. Since the temporal profile of the dressing laser pulse intensity is mapped onto $\varphi(\tau)$ via Eq. (4), it is possible to determine both the pulse duration and intensity.

fields and will greatly enhance the comparability of experiment and theory [16,17].

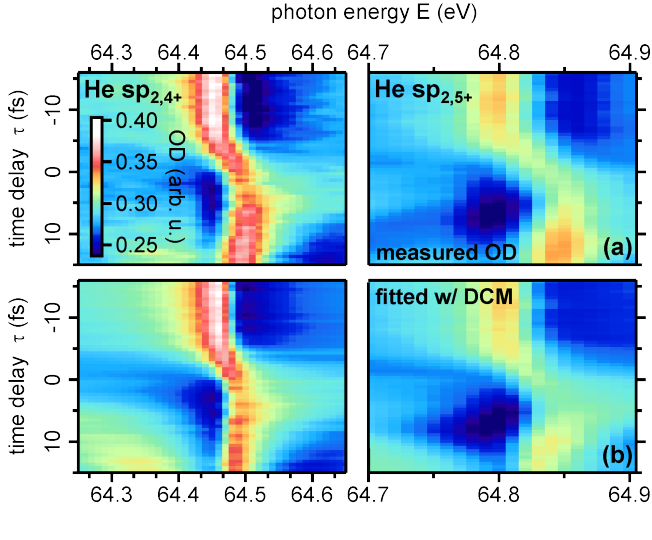 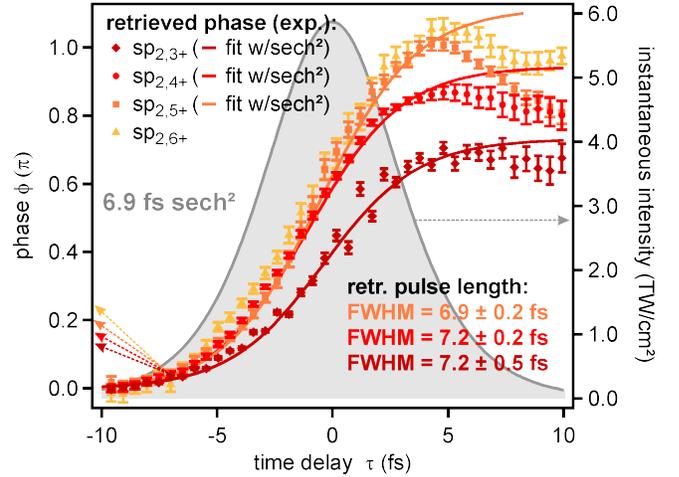

Fig. 2. Pulse characterization (experiment). (a) Measured time-delay dependent absorption spectra of the helium sp$_{2,4+}$ and sp$_{2,5+}$ Fano resonance lines dressed with a few-cycle NIR pulse that leads to a transient ponderomotive energy shift. In order to obtain better statistics, the data is binned along the $\tau$ axis for about one laser cycle (16 time-delay steps). (b) Result of fitting the experimental data with the DCM function from Eq. (5). (c) Laser-induced phase $\varphi(\tau)$ retrieved from the fit as a function of time delay. A fit with an integrated sech$^2$ profile (solid lines) yields the pulse duration and intensity. The pulse profile obtained this fit is plotted in gray with a duration of 6.9 fs full-width-at-half-maximum (FWHM) intensity.

The dipole response $d(t)$ of a freely decaying atomic state after excitation by a short Dirac-$\delta$-like pulse is expressed in the form

$$d(t > 0) \propto - i\, e^{i\omega_r t - \frac{\Gamma}{2}t + i\varphi_0}. \quad (1)$$

Here, $\omega_r$ denotes the state's resonance frequency, $\Gamma$ its decay rate, and the initial phase $\varphi_0$ accounts for the line shape [11]. It has been shown experimentally that this initial phase, and thus the absorption line shape, can be controlled by a femtosecond laser pulse interacting with the excited state directly after it has been populated. By assuming $\delta$-like excitation and equally short interaction, we have recently introduced a fully analytical dipole control model (DCM) for arbitrary time delay [15], which describes the experimental data well if the pulses are short compared to the lifetimes of the states. A similar approach was applied successfully in ref. [18] in order to describe strong field ionization of excited xenon levels. The basic principle of the DCM is depicted in Fig. 1 (a). We describe the perturbed polarization decay as

$$d_\tau(t,\tau) \propto \begin{cases} 0 & t < 0 \\ f_1(t) & 0 < t < \tau \\ A(\tau)\, f_1(t) & t > \tau \end{cases} \quad (2)$$

After the excitation by the XUV pulse at $t = 0$, the system evolves freely, in all generality given by the function $f_1(t)$, e.g. as in Eq. (1). Subsequent interaction with the NIR pulse perturbs the system, and accordingly the dipole response. After this interaction the temporal structure is again governed by the field-free evolution. Therefore, we can write the perturbed dipole response as the field-free response modified in amplitude and phase, which is parametrized by the complex quantity $A(\tau)$. For our pulse characterization method, we study the absorption response of highly excited atomic states. Dressed by the femtosecond laser field, the nearly free electron will be subject to a ponderomotive shift of its energy during the pulse. Such kind of transient energy shift was observed as a shift of the ionization threshold in helium [19]. Instead of studying the energy displacement, we focus on the transient phase shift that occurs while the atom is dressed, and which gives rise to a modified line shape. The additional phase thus acquired is given by

$$\varphi_P(t,\tau) = \int_0^t U_P(t' - \tau)\,dt' \quad (3)$$

Here, $U_P$ is the ponderomotive potential which is proportional to the laser intensity and the square of wavelength. In the limit of the DCM, the full phase shift $\varphi(\tau) = \varphi_P(t \to \infty, \tau)$ is treated to occur at the instance of interaction $\tau$. The phase jump is equal to the accumulated phase the atomic state experiences. Thus, we can write $A(\tau)$ in Eq. (2) as

$$A(\tau) = e^{i\varphi(\tau)} = \exp\left[i \int_0^\infty U_P(t - \tau)\,dt\right] \quad (4)$$

The approximation made by the DCM is illustrated in Fig. 1 (b). We then can analytically calculate the time-delay dependent spectrum $\tilde{D}_\tau$ of the dipole response:

$$\tilde{D}_\tau(\omega,\tau) \propto \begin{cases} \dfrac{1 - e^{i(\omega_r - \omega)\tau - \Gamma/2\,\tau}(1 - e^{i\varphi(\tau)})}{\omega_r - \omega + i\,\Gamma/2}\, e^{i\varphi_0} & \tau > 0 \\[6pt] \dfrac{e^{i\varphi(\tau)}}{\omega_r - \omega + i\,\Gamma/2}\, e^{i\varphi_0} & \tau < 0 \end{cases} \quad (5)$$

The imaginary part of Eq. (5) is proportional to the optical density (OD) of the target, and can thus be measured in ATAS.

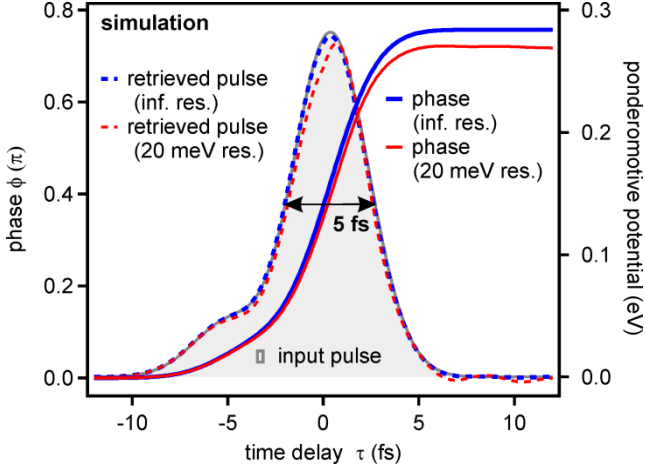

Fig. 3. Pulse characterization (simulation) for a validity check of the method. The input pulse (solid grey line and area) is compared to the pulse form (solid line) that is calculated from the retrieved phase (dashed line). Depicted are the results obtained for infinite (blue) and finite (red) spectrometer resolution. The XUV pulse used has a duration of 150 as FWHM.

In the following, we apply Eq. (5) to measured and simulated spectra in order to retrieve the laser induced phase $\varphi(\tau)$.

In the experiment, we use near-infrared laser pulses of 30 fs duration and 0.7 mJ pulse energy, which are spectrally broadened in a neon-filled hollow-core fiber and compressed with chirped mirrors. The beam is then focused into a gas cell and drives high-order harmonic generation to provide phase-locked attosecond-pulsed XUV light. Subsequently, the NIR intensity is controlled by an aperture, and a time delay between the driving laser pulse and the XUV pulses is introduced by means of a piezo-driven split mirror. Both beams are focused into the sample gas cell and the transmitted XUV light is analyzed in a home-built spectrometer with 20 meV (s.d.) resolution.

Figure 2 (a) shows measured time-delay dependent absorption spectra of the helium $sp_{2,4+}$ and $sp_{2,5+}$ doubly excited states excited by attosecond-pulsed XUV light. With energies above 64.4 eV these states are close to the $2s^1/2p^1$-ionization threshold of helium, and are therefore strongly affected by the NIR dressing laser. The lifetime of the $sp_{24+}$ state, is about 190 fs [20], which is much longer than the expected pulse duration, and renders the DCM applicable to the studied system. For $\tau < -10$ fs the NIR pulse does not act on the states and the line profile corresponds to the freely decaying dipole with a natural Fano line shape [21]. For $\tau \gg 0$, only a small fraction of the decaying dipole response is affected by the NIR pulse and contributes to the reshaping of the lines, converging to the original lines for $\tau \to \infty$. In the vicinity of the pulse overlap, the absorption lines are modified by the dressing laser. Therefore, this temporal region is used to reconstruct the laser pulse.

To determine the laser-imprinted phase $\varphi(\tau)$, we fit the imaginary part of Eq. (5) to each lineout of Fig. 2 (a). The transition frequency and the initial phase are determined for the unperturbed line, and are fixed in the analysis.

The fitted spectra are shown in Fig. 2 (b) and the retrieved phase $\varphi(\tau)$ is displayed in Fig. 2 (c). In addition, the results obtained for the $sp_{2,3+}$ and $sp_{2,6+}$ states are also depicted. Except, for the $sp_{2,3+}$ state, we do not observe signatures of resonant coupling, e.g. a sub-cycle oscillation of the optical density. Thus, we can assume a purely non-resonant coupling scenario, which leads for the highest lying states to a ponderomotive shift of the energy. The $sp_{2,3+}$ is resonantly coupled to the 2s2p state, however, the corresponding beating structure cancels due to averaging over one optical cycle such that only the non-resonant phase shift remains. The figure reveals that the accumulated phase is increasing with the principle quantum number $n$, and approaching each other for large $n$. This is in agreement with the intuitive expectation that the bound states closest to the continuum are more strongly affected by the laser field. The method allows us to quantify this finding: The action, i.e. the accumulated phase, which the electron experiences in the $sp_{2,6+}$ state is roughly 60 % larger than the action which the electron experiences in the $sp_{2,3+}$ state.

The phase maximum is reached after a delay $\tau \approx 5$ fs. For larger time delays, and especially for the highly excited states, the retrieved phase starts to decline for yet unknown reasons, thus we will restrict the data range used for the laser pulse measurement to $\tau < 5$ fs. A possible explanation is the close proximity of the higher-lying resonances, which causes a slight $\tau$-dependent overlap of the lines due the limited resolving power of our spectrometer. In this case, the method would directly benefit from a higher spectrometer resolution. Also the sub-picosecond pedestal of the laser pulse can affect the atomic line shape. A laser field with duration on the order of the +100 fs lifetime of the studied states is currently not considered in our reconstruction model, which assumes a strong-field pulse much shorter than the states' lifetime.

We now focus on the three states that exhibit the strongest absorption signal and are spectrally best resolved, i.e. $sp_{2,3+}$, $sp_{2,4+}$ and $sp_{2,5+}$, which allow for the most reliable determination of the phase $\varphi(\tau)$, as well as for a quantitative retrieval of the laser-pulse characteristics. This is confirmed by the good agreement with the weaker $sp_{2,6+}$ state. We can reliably fit $\varphi(\tau)$ of the states with an integrated sech$^2$ temporal profile, which fits the data better than the standard Gaussian and will be referred to as $\varphi_{\text{fit}}(\tau)$. This corresponds to a laser pulse of $6.9 \pm 0.2$ fs FWHM intensity duration for the $sp_{2,5+}$ and similar values for the other two states. This shows that the determination of the pulse duration is relatively insensitive to the choice of the state and hence the assumed limit of a pure ponderomotive shift.

This situation is different for the characterization of the absolute pulse intensity, because here the absolute value of the laser-imprinted phase enters. As we know from Eq. (4), $\varphi(\tau)$ is the integrated ponderomotive potential of the laser pulse if the electron is close to free. Hence, the derivative $d\varphi(\tau)/d\tau$ for a loosely bound state directly yields the time-dependent ponderomotive energy, or by knowing the center frequency of the pulse, the time-resolved laser intensity. Using the fit of the sp$_{2,5+}$ curve for the evaluation yields $\hbar\, d\varphi_{\text{fit}}(\tau)/d\tau = 0.30$ eV at the peak of the NIR pulse. To account for the fact that the phase imprinted on the most-weakly bound state which we can analyze, i.e. the sp$_{2,6+}$, is about 5 % higher compared to the sp$_{2,5+}$ we scale the phase derivative by this factor. With that the peak ponderomotive energy amounts to $U_{\text{P}} = 0.32$ eV. For the center wavelength of 730 nm, this corresponds to a peak intensity of $6.4 \times 10^{12}$ W/cm$^2$. The exact value depends on the validity of the assumption $d\varphi(\tau)/d\tau = U_{\text{P}}$. Therefore, the method can be improved by incorporating a more accurate description of the laser-dressed states. The intensity sensitivity for the depicted measurement is about $0.5 \times 10^{12}$ W/cm$^2$, which could be improved by reducing measurement noise, *e.g.* via an increased pulse-to-pulse stability of the laser system. As both XUV pulse length and delay step size can be <100 as, even the characterization of single-cycle pulses should be experimentally feasible.

In order to validate the applicability of the retrieval method, in particular the capability of the DCM to describe transient phenomena, we simulated time-delay scans of an isolated state exposed to the ponderomotive potential of a dressing few-cycle laser pulse. The population of the excited state via the XUV pulse is treated perturbatively, and the phase evolution of the state is continuously modified according to Eq. 3. We used values for $\omega_{\text{r}}$, $\varphi_0$, $\Gamma$ and $U_{\text{P}}$ similar to the experimental conditions. In order to simulate the finite spectrometer resolution, additional calculations were performed where the natural line is convolved with a 20 meV Gaussian filter function. Figure 3 shows the laser imprinted phase $\varphi(\tau)$ retrieved by the fit and the corresponding pulse shape, as well as the input pulse form, which we chose to be a 5 fs FWHM Gaussian pulse with a weak post pulse. The pulse shape was directly obtained using the derivative $d\varphi(\tau)/d\tau$. Hence, we need no specific assumption about the input pulse, except that it is short compared to the state's lifetime. For the fully resolved spectral line shape, the agreement of retrieved and input pulse is excellent, whereas the convolution with the spectrometer response introduces a slight deviation for the reconstructed intensity on the order of 3 %. The influence of the XUV pulse duration on the retrieved NIR pulse duration is small: going from $\delta$-like excitation to a 2000 as pulse, which covers most experimental situations, introduces a deviation of about 5 %.

To summarize, we introduced a method that allows for the measurement of both the duration and intensity of few-cycle laser pulses using an ATAS setup. The strong-field NIR laser-pulse parameters are determined *in situ,* i.e. at the location where the investigated sample interacts with the intersecting NIR and XUV pulses – or in other words – where the physics of quantum-dynamics experiments happens. The results can thus be directly used for data analysis, combining measurement and pulse characterization in one single experiment. Future applications of the method will thus enable precision measurements of dynamical phase shifts in response to well characterized laser fields, leading to a better understanding of bound-state non-perturbative light-matter interaction on ultrafast time scales.

**References**

1. R. Kienberger, E. Goulielmakis, M. Uiberacker, A. Baltuska, V. Yakovlev, F. Bammer, A. Scrinzi, Th. Westerwalbesloh, U. Kleineberg, U. Heinzmann, M. Drescher, and F. Krausz, *Nature* **427**, 817-821 (2004)
2. E. Goulielmakis, M. Uiberacker, R. Kienberger, A. Baltuska, V. Yakovlev, A. Scrinzi, Th. Westerwalbesloh, U. Kleineberg, U. Heinzmann, M. Drescher, and F. Krausz, *Science* **305**, 1267-1269 (2004)
3. E. Goulielmakis, V. S. Yakovlev, A. L. Cavalieri, M. Uiberacker, V. Pervak, A. Apolonski, R. Kienberger, U. Kleineberg, and F. Krausz, *Science* **317**, 769-775 (2007)
4. Z.-H. Loh, M. Khalil, R. E. Correa, R. Santra, C. Buth, and S. R. Leone, *Phys. Rev. Lett.* **98**, 143601 (2007)
5. H. Wang, M. Chini, S. Chen, C.-H. Zhang, F. He, Y. Cheng, Y. Wu, U. Thumm, and Z. Chang, *Phys. Rev. Lett.* **105**, 143002 (2010)
6. E. Goulielmakis, Z.-H. Loh, A. Wirth, R. Santra, N. Rohringer, V. S. Yakovlev, S. Zherebtsov, T. Pfeifer, A. M. Azzeer, M. F. Kling, S. R. Leone, and F. Krausz, *Nature* **466**, 739-743 (2010)
7. R. Santra, V. S. Yakovlev, T. Pfeifer, and Z.-H. Loh, Phys. Rev. A **83**, 033405 (2011)
8. M. B. Gaarde, C. Buth, J. L. Tate, and K. J. Schafer, *Phys. Rev. A* **83**, 013419 (2011)
9. M. Holler, F. Schapper, L. Gallmann, and U. Keller, *Phys. Rev. Lett.* **106**, 123601 (2011)
10. M. Schultze, E. M. Bothschafter, A. Sommer, S. Holzner, W. Schweinberger, M. Fiess, M. Hofstetter, R. Kienberger, V. Apalkov, V. S. Yakovlev, M. I. Stockman, and F. Krausz, *Nature* **493**, 75-78 (2013)
11. C. Ott, A. Kaldun, P. Raith, K. Meyer, M. Laux, J. Evers, C. H. Keitel, C. H. Greene, and T. Pfeifer, *Science* **340**, 716-720 (2013)
12. C. Ott, A. Kaldun, L. Argenti, P. Raith, K. Meyer, M. Laux, Y. Zhang, A. Blättermann, S. Hagstotz, T. Ding, R. Heck, J. Madronero, F. Martin, and T. Pfeifer, *Nature* **516**, 374-378 (2014)
13. Lin, C. D., and Chu, W.-C. Controlling Atomic Line Shapes. *Science* **340**, 694-695 (2013)
14. A. Kaldun, C. Ott, A. Blättermann, M. Laux, K. Meyer, T. Ding, A. Fischer, and T. Pfeifer, *Phys. Rev. Lett.* **112**, 103001 (2014)
15. A. Blättermann, C. Ott, A. Kaldun, T. Ding, and T. Pfeifer, *J. Phys. B: At. Mol. Opt. Phys.* **47**, 124008 (2014)
16. M. B. Gaarde, C. Buth, J. L. Tate, and K. J. Schafer, Phys. Rev. A **83**, 013419 (2011)
17. R. Santra, V. S. Yakovlev, T. Pfeifer, and Z.-H. Loh, Phys. Rev. A **83** 033405 (2011)
18. B. Bernhardt, A. R. Beck, X. Li, E. R. Warrick, M. J. Bell, D. J. Haxton, C. W. McCurdy, D. M. Neumark, and S. R. Leone, *Phys. Rev. A* **89**, 023408 (2014)
19. M. Chini, X. Wang, Y. Cheng, Y. Wu, D. Zhao, D. A. Telnov, S.-I. Chu, and Z. Chang, *Sci. Rep.* **3**, 1105 (2013)
20. K. Schulz, G. Kaindl, M. Domke, J. D. Bozek, P. A. Heimann, A. S. Schlachter, and J. M. Rost, *Phys. Rev. Lett.* **77**, 3086 (1996)
21. U. Fano, *Phys. Rev.* **124**, 1866-1878 (1961)

The authors acknowledge financial support from the Deutsche Forschungsgemeinschaft (grant PF 790/1-1) and the European Reseach Concil (ERC, grant X-MuSIC-616783).